\documentclass[a4paper,11pt]{article}

\usepackage[norefcommands]{heppub}
\usepackage{bm}
\usepackage{graphicx}
\usepackage{wrapfig}
\usepackage{slashed}
\usepackage[normalem]{ulem}
\usepackage[utf8]{inputenc}
\usepackage{mathtools}
\usepackage{stmaryrd}
\usepackage{appendix}
\usepackage{xspace}
\usepackage{upgreek}
\usepackage{multirow}

\usepackage{scrextend}

\usepackage{tabularx}
\usepackage{tensor}

\usepackage{marginnote}

\definecolor{darkgreen}{rgb}{0.0, 0.4, 0.0}

\newcommand{\df}{\mathrm{d}}

\newcommand{\as}{\alpha_s}

\newcommand{\Pythiaxx}{\texttt{Pythia\xspace8.3}\xspace}
\newcommand{\Pythia}{\texttt{Pythia}\xspace}

\newcommand{\Vinciaxx}{\texttt{Vincia\xspace2.3}\xspace}
\newcommand{\Herwig}{\texttt{Herwig}\xspace}
\newcommand{\Herwigxx}{\texttt{Herwig\xspace7.3}\xspace}

\newcommand{\Fastjet}{\texttt{FastJet}\xspace}

\def\df{\textrm{d}}

\makeatletter
		
		\newcommand{\AfterMySpacing}{%
			\vskip\dp\strutbox
			\prevdepth\dp\strutbox
			\noindent\strut}  

\makeatother

\DeclareRobustCommand{\Refcite}[1]{Ref.~\cite{#1}}

\DeclareRobustCommand{\eq}[1]{eq.~\eqref{eq:#1}}

\DeclareRobustCommand{\secn}[1]{\hyperref[sec:#1]{section~\ref*{sec:#1}}}
\DeclareRobustCommand{\Sec}[1]{\hyperref[sec:#1]{Section~\ref*{sec:#1}}}

\DeclareRobustCommand{\subsec}[1]{\hyperref[subsec:#1]{subsection~\ref*{subsec:#1}}}
\DeclareRobustCommand{\Subsec}[1]{\hyperref[subsec:#1]{Subsection~\ref*{subsec:#1}}}

\DeclareRobustCommand{\app}[1]{\hyperref[app:#1]{appendix~\ref*{app:#1}}}
\DeclareRobustCommand{\App}[1]{\hyperref[app:#1]{Appendix~\ref*{app:#1}}}

\DeclareRobustCommand{\fig}[1]{\hyperref[fig:#1]{figure~\ref*{fig:#1}}}
\DeclareRobustCommand{\Fig}[1]{\hyperref[fig:#1]{Figure~\ref*{fig:#1}}}

\DeclareRobustCommand{\tab}[1]{\hyperref[tab:#1]{table~\ref*{tab:#1}}}
\DeclareRobustCommand{\Tab}[1]{\hyperref[tab:#1]{Table~\ref*{tab:#1}}}

\usepackage{xurl}

\preprint{\begin{flushright}
DESY-24-107\\
UWThPh 2024-14
\end{flushright}}

\title{Top Quark Mass Extractions from Energy Correlators: A Feasibility Study}

\author[a]{Jack Holguin,}
\affiliation[a]{Consortium for Fundamental Physics, School of Physics \& Astronomy, \\
University of Manchester, Manchester M13 9PL, United Kingdom}

\author[b]{Ian Moult,}
\affiliation[b]{Department of Physics, Yale University, New Haven, CT 06511}

\author[c]{Aditya Pathak,}
\affiliation[c]{Deutsches Elektronen-Synchrotron DESY, Notkestr. 85, 22607 Hamburg, Germany}

\author[d]{Massimiliano Procura,}
\affiliation[d]{University of Vienna, Faculty of Physics, Boltzmanngasse 5, A-1090 Vienna, Austria}

\author[e]{Robert Sch\"ofbeck,}
\author[e]{Dennis Schwarz}
\affiliation[e]{Institute for High Energy Physics, Austrian Academy of Sciences, Dominikanerbastei 16, A-1010 Vienna, Austria}

\emailAdd{jack.holguin@manchester.ac.uk}
\emailAdd{ian.moult@yale.edu}
\emailAdd{aditya.pathak@desy.de}
\emailAdd{mprocura@univie.ac.at}
\emailAdd{robert.schoefbeck@oeaw.ac.at}
\emailAdd{dennis.schwarz@cern.ch}

\abstract{In a recent article, we proposed an energy correlator-based method to achieve a precision top quark mass extraction from jet substructure, using the $W$-boson mass as a standard candle.
In this paper, we perform an extensive event generator simulation study of this proposal, testing its experimental viability and its sensitivity to different subprocesses in the top quark production and decay.
On the experimental side, we show that uncertainties in the jet energy scale, constituent energy scale, and tracking efficiency have a minimal effect.
On the theoretical side, we find that our observable isolates the perturbative decay of the top quark. At the same time, nonperturbative physics, such as the modelling of color reconnection and the underlying event, has a negligible impact on the distribution.
We conclude that our proposed measurement is resilient to the experimental and theoretical aspects of the hadron collider environment, with variations in model parameters consistently leading to $\lesssim 100~$MeV shifts in the measured top mass.
Our results motivate precision theoretical calculations of the energy correlator on top decays, both analytic and using parton shower generators, and further exploration of the experimental measurement.}


\begin{document}

\maketitle

\section{Introduction}
\label{sec:intro}

A precision extraction of the top quark mass at hadron colliders is a notoriously complex task, both experimentally and theoretically. Numerous measurements of the top quark mass have been performed~\cite{CDF:2014upy,CMS:2015lbj,ATLAS:2016muw,ParticleDataGroup:2020ssz}, but uncertainties in their theoretical interpretation still limit their significance as crucial high-precision tests of the Standard Model. For reviews, see \cite{Nason:2017cxd,Hoang:2020iah}.

Energy correlators \cite{Basham:1978bw,Basham:1977iq,Ellis:1978ty,Basham:1979gh,Basham:1978zq} have recently been made into a practical jet substructure observable in hadron colliders \cite{Dixon:2019uzg,Chen:2020vvp,Komiske:2022enw}, capable of identifying both scaling \cite{Chen:2020vvp,Lee:2022ige,Chen:2023zlx} and shape \cite{Chen:2019bpb,Chen:2020adz,Chen:2022swd,Chen:2022jhb} information about the distribution of energy inside jets. They have been measured in proton-proton collisions by the CMS~\cite{CMS:2024mlf}, RHIC~\cite{Tamis:2023guc}, and ALICE~\cite{Fan:2023} experiments, and in heavy ion collisions by CMS \cite{CMS-PAS-HIN-23-004}, illustrating their experimental feasibility. Indeed, they have already led to the most precise extraction of the strong coupling constant, $\alpha_s$, from jet substructure~\cite{CMS:2024mlf}. Recent work, \Refcite{Xiao:2024rol}, has highlighted that energy correlator measurements on top jets have several experimental advantages relative to commonly used jet observables.

In this context, it has been proposed \cite{Holguin:2022epo,Holguin:2023bjf} that the three-point energy correlator, measured on sufficiently boosted top quark jets, can be used for a precision extraction of the top quark mass. Since energy correlators allow for the determination of dimensionless characteristic angles, as opposed to dimensionful mass scales, a clean approach to extracting the top quark mass in a well-defined mass scheme from LHC data is to isolate the ratio of the top to the $W$ boson mass from selected projections of the correlator spectrum~\cite{Holguin:2023bjf}. The goal of this paper is to thoroughly explore, using event generator simulations, the robustness of this proposal for the precision extraction of a theoretically well-defined top mass \cite{Hoang:2009yr,Hoang:2017suc} and to motivate further theoretical and experimental research into energy correlators on top quark decays. A preliminary version of the results of this paper was presented in \Refcite{AdiCMSTopPAG}.

\begin{figure}
\centering
\includegraphics[width=\textwidth]{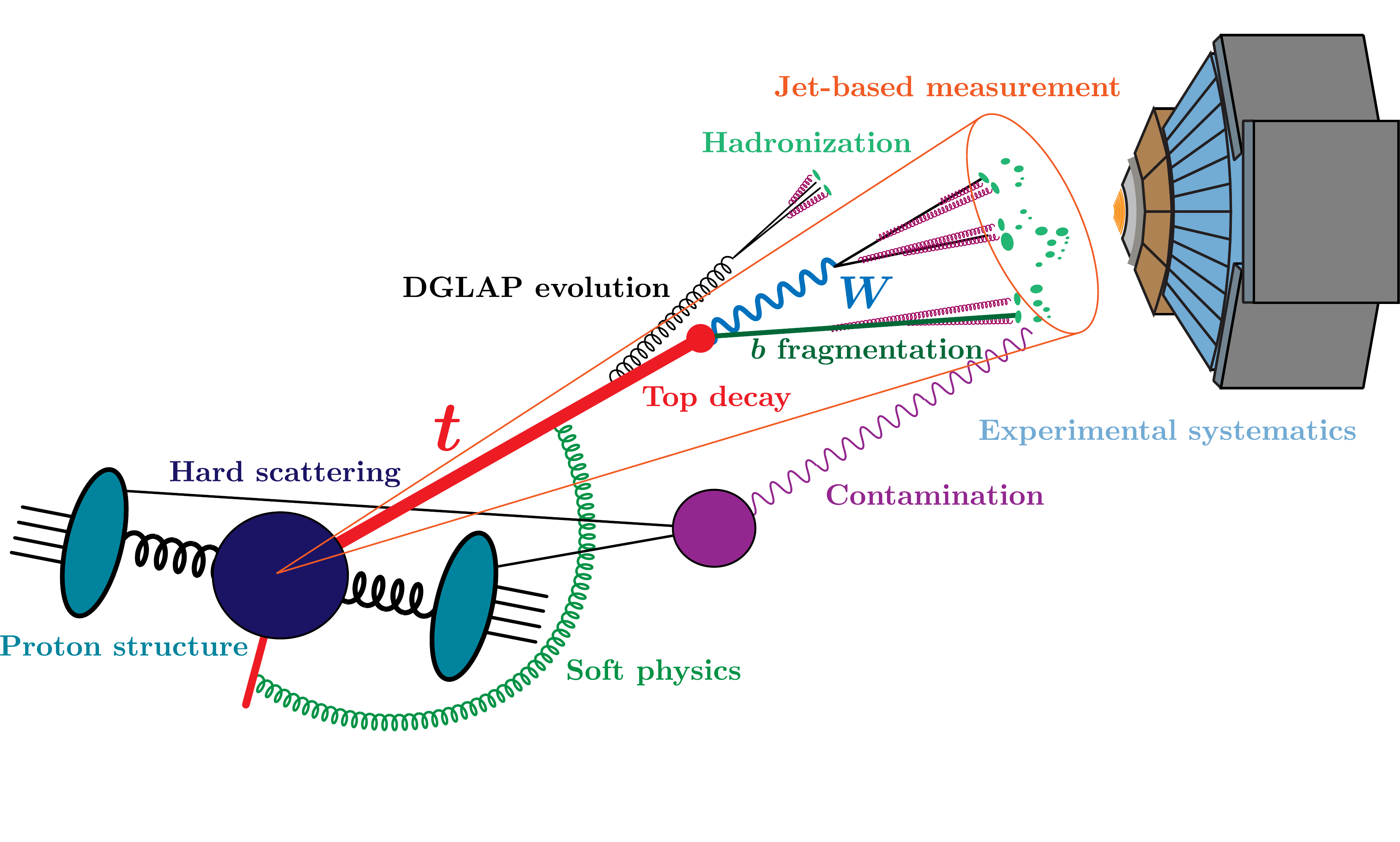}
\caption{An illustration of the many subprocesses that enter into the description of a top quark jet substructure observable at the LHC, measured by the detector with varying degrees of angular and energy resolution.}
\label{fig:fact}
\end{figure}

The precise extraction of the top quark mass from a measurement of energy correlators on top quark jets relies on a detailed understanding of the numerous physics subprocesses involved in the production and decay of a boosted top quark, as well as the experimental systematics, as illustrated in \fig{fact}. To show the feasibility of the measurement, we must show that both the experimental systematics and the dependence on poorly understood nonperturbative subprocess are under control. We do this by systematically investigating the dependence of the proposed measurement on each of the subprocesses in \fig{fact} using event generator simulations.

On the experimental side, we show in simulations that, being primarily an angular measurement, the top quark mass extracted with our method remains resilient to uncertainties in the jet energy scale, the constituent energy scale, and the simulated particle tracking efficiency. We find that these effects for top jet transverse momenta $p_{T,\rm jet}$ larger than $400~$GeV have an impact of $\mathcal{O}(100)$ MeV on the extracted top quark mass.

On the theoretical side, we show that our method achieves remarkable robustness against the modelling of the nonperturbative subprocess, including hadronization, color reconnection, underlying event (UE), and variations in PDF sets, with each impacting the extracted top quark mass by less than $\mathcal{O}(100)$ MeV. Additionally, the observable shows high resilience to perturbative subprocesses occurring outside the identified top quark jet and the details of the top quark production mechanism, which we probe by comparing LO and NLO hard matching. This provides strong evidence that the observable has effectively isolated the dynamics associated with the top quark decay. In this context, we indeed observe an important dependence on the top-jet dynamics, particularly on the perturbative description of the top decay and the jet radius. We find that differences between the approximation schemes in the modelling of the top decay can lead to shifts in the extracted top mass as large as $1~$GeV. This strongly motivates further steps on the analytic side that remove the dependence on event generators to achieve the target precision and release public event generator simulations with complete NLO top decays for insightful comparisons. In addition, the small dependence on event generator uncertainties is promising from the perspective of performing a complete unfolded measurement. Since the exact impacts of the resulting uncertainties in the unfolding strongly depend on the setup, these are not subject to this study.

An outline of this paper is as follows. In \secn{review}, we review our proposal for precision measurement of the top quark mass with energy correlators on boosted top quark jets. \Sec{exp} contains a systematic study of experimental uncertainties, in particular, those associated with the jet energy scale, constituent energy scale, and tracking efficiency based on parametrizations of the CMS detector response \cite{CMS:2019csb,CMS:2024irj}. \Sec{EventGen} is devoted to the study of the uncertainties in event generator modelling and groups effects by whether analytical understanding from a perturbative calculation, beyond event generator simulations, could be achieved within the factorized framework outlined in \Refcite{Holguin:2023bjf}. We conclude in \secn{conc}.

\section{Top Quark Mass Measurements with Energy Correlators}
\label{sec:review}

\begin{figure}
\centering
\includegraphics[width=0.65\textwidth]{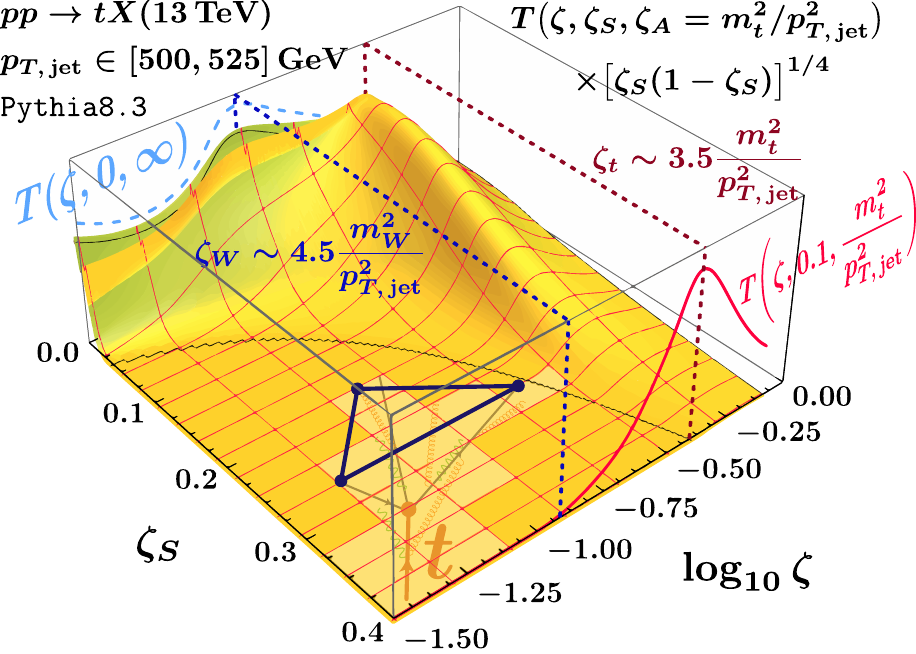}
\caption{An illustrative figure produced from \Pythia showing the imprint of top and $W$ on the 3-point EEC. First presented in \Refcite{Holguin:2023bjf}.}
\label{fig:armadillo}
\end{figure}

In this section, we review our detailed proposal of \Refcite{Holguin:2023bjf} to measure the top quark mass in the jet substructure using energy correlators.
In short, we propose to extract the top quark mass from a simultaneous fit of two features in the spectrum of the three-point energy correlator. These are shown in \fig{armadillo} as a prominent ridge which imprints the top decay (projected on the front face for a specific value of $\zeta_S$) and a second smaller angle peak region formed from the boosted $W$ decay (projected on the back face with the asymmetry cut $\zeta_A$ removed). To facilitate these fits, we introduce two projections of the 3-point correlator, $T$ and $W$, each isolating the relevant features of the associated decay.

We define the top projection ($T$)\footnote{ Regarding the experimental measurement, we note that the distribution $T$ (as well as $W$ defined below) and the hadronic cross-sections in the all present formulae are differential in $p_{T, {\rm jet}}$ and should be integrated over the narrow $p_{T, {\rm jet}}$ bins in which the measurements are performed. The differentials have been omitted for brevity.} as
\begin{align}
T(\zeta, \zeta_S, \zeta_{A};m_{W},m_{t},p_{T, {\rm jet}}) \equiv & \sum_{{}^{\mathrm{hadrons}}_{ i,j,k \in {\rm jet}}}\int \df \zeta_{ijk} ~
\frac{\df \sigma_{i,j,k}}{\df
\zeta_{ijk}} ~ \frac{p_{T,i} \, p_{T,j} \, p_{T,k}}{\big(p_{T,\mathrm{jet}}\big)^{3}} ~ \delta\left(\zeta -
\frac{(\sqrt{\zeta_{ij}}+\sqrt{\zeta_{jk}})^2}{2} \right) \nonumber \\[-4pt]
&\Theta(\zeta_{ij}\geq\zeta_{jk}\geq\zeta_{ki}\geq \zeta_{S}) ~ \Theta\left(\zeta_{A} >(\sqrt{\zeta_{ij}}-\sqrt{\zeta_{jk}})^2 \right)\,. \label{eq:observable}
\end{align}
Here, the sum extends over all (not necessarily distinct) triplets of hadrons $(i,j,k)$ in the most energetic jets within each hemisphere of an identified top event. In \eq{observable}, $p_{T,i}$ is the transverse momentum of particle $i$, which provides the energy weighting, and $\df \zeta_{ijk} \equiv \df \zeta_{ij} \df \zeta_{jk} \df \zeta_{ki}$. The quantity $T$ is measured as a function of $\zeta$ while $\zeta_S$ and $\zeta_A$ are used to enforce cuts on the measurement, revealing the top decay feature. In particular, $\zeta_S$ is the smallest allowed angle in the correlator, and $\zeta_A$ is the maximum allowed asymmetry between the two longest sides of the triangular configuration. In this work, $\zeta_S$ and $\zeta_A$ are fixed at the particular values $\zeta_S= 0.8 (172 [\mathrm{GeV}]/p_{T \, \mathrm{jet}})^2$ and $\zeta_A= (172 [\mathrm{GeV}]/p_{T \, \mathrm{jet}})^2$ to cleanly isolate the top peak feature.

To isolate the $W$-imprint, we define a projection of the 3-point correlator and take the ratio of this projection to the 2-point correlator. The ratio is introduced to reduce the effect of nonperturbative physics in the distribution and is defined as
\begin{align}
\label{eq:ratio}
&W(\zeta;M_{W},M_{t},p_{T, {\rm jet}}) \nonumber \\[-10pt] &~~~~~\equiv T(\zeta,0,\infty;m_{W},m_{t},p_{T, {\rm jet}}) \left(\sum_{{}^{\mathrm{hadrons}}_{~~ i,j}}\int \mathrm{d} \zeta_{ij} ~ \frac{p_{T,i} \,
p_{T,j}}{\big(p_{T,\mathrm{jet}}\big)^{2}} ~ \frac{\mathrm{d} \sigma_{i,j}}{\mathrm{d} \zeta_{ij}}~ \delta(\zeta -
\zeta_{ij} )\right)^{-1},
\end{align}
where the denominator is the standard two-point EEC, while the numerator, $T(\zeta,0,\infty)$, is the correlator of \eq{observable}, with no cuts on the asymmetry $\zeta_A$, or smallest angle, $\zeta_S$.
The ultimate goal for extracting the top quark mass in this approach is the simultaneous fit of analytic calculations to the experimental measurements of the $T$ and $W$ distributions obtained from a sample of high-$p_{T}$ top jets. Using $m_W$ as input, the system is over-constrained, allowing for a unique determination of $m_t$, without dependence on $p_{T, {\rm jet}}$.

Since this paper aims to motivate a complete theoretical calculation and assess the accompanying systematic uncertainties, we use a simplified procedure that is amenable to testing analytic or simulation-based top mass fits of the $T$ and $W$ distributions. It consists of a fit of the peak positions (denoted by $\zeta_t$ and $\zeta_W$) in the two distributions,
\begin{align}
\frac{{\rm d} T}{{\rm d} \zeta}\Bigg|_{\zeta = \zeta_t} = 0, ~~~~~ \frac{{\rm d} W}{{\rm d} \zeta}\Bigg|_{\zeta = \zeta_W} = 0. \label{eq:HowToDifferentiate}
\end{align}
These equations lead to a completely constrained system in terms of $m_t$, $p_{T, {\rm jet}}$\footnote{We stress that, for fixed $p_{T, {\rm jet}}$, these equations are fully constrained but not over-constrained in terms of $m_t$, $p_{T, {\rm jet}}$, as would be the case if the entire $W$ and $T$ functions are simultaneously fitted.} enabling a top mass determination while having the added benefit that they can be solved easily with event generator simulations and/or with data samples.
The simulation-based assessment of the various uncertainties in the peak-finding-based pseudo measurement, described in the following sections, is a conservative estimate of the uncertainties directly using the theoretical predictions.
However, it is important to stress that a complete fit from theoretical predictions of $T$ and $W$ to the experimental data will always perform at least as well and likely better than this simple approach.

In the large boost limit of a top quark decay, we have the relation
\begin{align}
m_{t}= m_{W} \left[ C(\as,R) \sqrt{\zeta_{t}/\zeta_W} + \mathcal{O}\left(\frac{m_{W}}{ p_{T,{\rm jet}} }, \frac{m_{t}}{ p_{T,{\rm jet}}} \right)\right] , \label{eq:C}
\end{align}
where the coefficient $C$, which has a perturbative expansion and depends on the jet radius $R$, characterizes the relationship between the top decay angle and the $W$ decay angle. This relationship is determined by the energy-weighted boost received by the $W$ in the rest frame of the top decay, averaged across events, and is a necessary input from theory found by solving eqs.~\eqref{eq:HowToDifferentiate}. For the present work, we extract $C$ by inverting \eq{C} with $\zeta_{t}$ and $\zeta_W$ extracted from parton level simulations of top jets with $m_t = 172.5~$GeV and $m_W = 80.4$~GeV. These parton-level simulations provide a proxy for naive, purely perturbative theoretical calculations of $T$ and $W$. The values we find are shown in \tab{Cvalues}. We stress that different event generators are expected to return slightly different values for $C$ since each generator uses different approximations to the NLO top decay. Furthermore, we stress that a self-consistent analysis with an event generator should use the value of $C$ returned by that generator, as is the case in the analyses below. We will discuss this in more detail in \secn{varyingpertmodels}.

\begin{table}
\centering
\begin{tabular}{|c||c|c|c|c|}
\hline Shower & $R$ = 0.8 & $R$ = 1 & $R$ = 1.2& $R$ = 1.5 \\ \hline\hline
\Pythiaxx & 1.075 $\pm$ 0.001 & 1.090 $\pm$ 0.001 & 1.099 $\pm$ 0.001 & 1.105 $\pm$ 0.001\\ \hline \Vinciaxx & 1.078 $\pm$ 0.001 & 1.091 $\pm$ 0.002 & 1.101 $\pm$ 0.001 & 1.107 $\pm$ 0.001\\ \hline \Herwigxx Dipole & 1.078 $\pm$ 0.001 & 1.088 $\pm$ 0.001 & 1.098 $\pm$ 0.001 & 1.106 $\pm$ 0.001\\ \hline \Herwigxx A.O. & 1.092 $\pm$ 0.001 & 1.104 $\pm$ 0.001 & 1.113 $\pm$ 0.001 & 1.120 $\pm$ 0.001\\ \hline
\end{tabular}
\caption{Values for the perturbative constant $C$ as extracted from parton-shower generators at parton level, for different values of the jet radii. These constants can be determined by applying perturbation theory to the top decay (largely at fixed order with only small effects from resummation) and are here extracted from the event generators in lieu of input from analytical calculations.}
\label{tab:Cvalues}
\end{table}

\subsection{Strategy for Peak Finding with Event Generator Data}

\begin{figure}[t]
\centering
\includegraphics[width=0.49\textwidth]{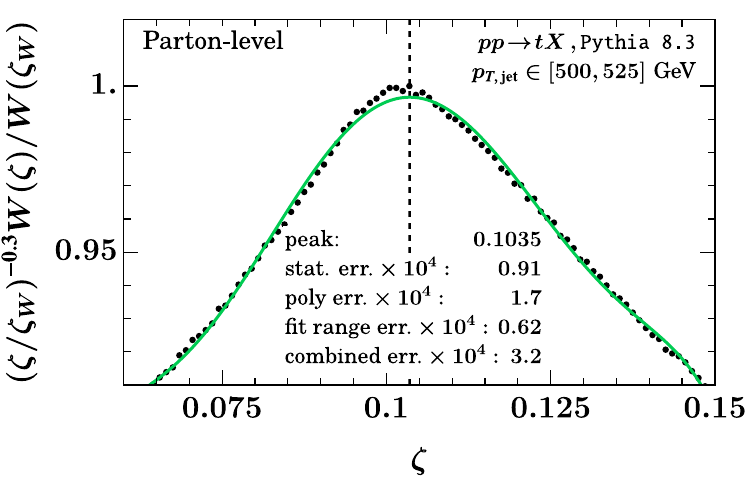}
\includegraphics[width=0.49\textwidth]{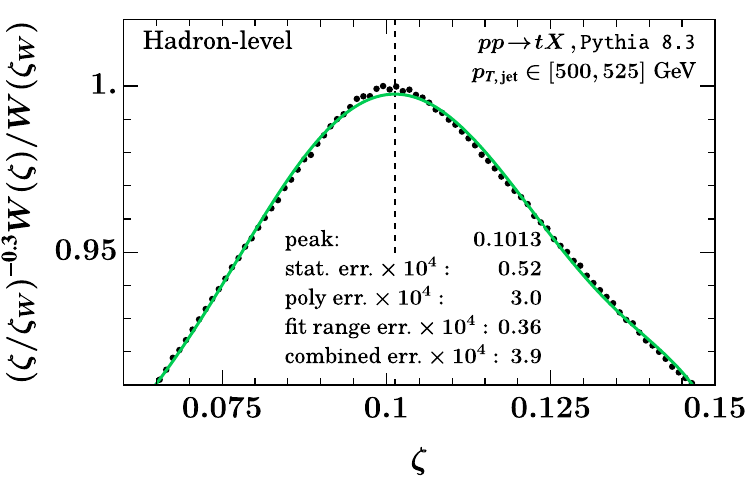}
\caption{Representative fits for peak locations in the $W$ distribution, shown here for \Pythiaxx, $p_T\in[500,525]$ GeV for hadron and parton level. The inset shows the uncertainty breakdown, with the dominant uncertainty arising from the variations in the polynomial order.}
\label{fig:Wfits}
\end{figure}

Without further input from analytical calculations, $\zeta_t$ and $\zeta_{W}$ must be found from fits of the peak regions in the $T$ and $W$ distributions. All fits in this paper were performed using fifth-degree polynomials for $\zeta_t$ and sixth-degree for $\zeta_W$. Sufficiently high number of Monte Carlo events were generated such that the statistical error is subdominant in the analysis below. Several fitting procedures were explored, such as Gaussian, crystal ball, and Breit-Wigner functions, and while consistent results were found, simple polynomial fits proved to be the most numerically stable. For the case of the $W$ distribution more care is required. While the hadron-level $W$ distributions displayed a smooth peaked feature for a range of $p_{T,\rm jet}$ bins, the parton level distributions displayed a small oscillating skew in a localised region around the peak. As a result, the parton-level peak position retains some sensitivity to the choice of degree of polynomial and the fit range. To this end, we restricted to a region roughly defined by half peak maximum and avoided zooming too narrowly around the peak. We discarded polynomials up to degree-five for $W$ distributions and found that degree-six and degree-seven polynomials resulted in equivalent fits. To account for ambiguity associated in the fitting procedure, we incorporate in our uncertainty the differences in the peak positions found via degree-six and degree-seven polynomials, and due to variation of the fitting window in the spectrum by $\pm 10\%$. Examples of the polynomial fits and a break down of the errors can be seen in Fig.~\ref{fig:Wfits}.

\section{Experimental Reconstruction Uncertainties}
\label{sec:exp}

In this section, we analyze the effects of uncertainties characterizing the experimental environment on our energy-correlator-based top mass extraction. Throughout this paper we use the \Herwigxx~\cite{Bahr:2008pv,Bahr:2008tf,Bellm:2019zci,Bewick:2023tfi}\footnote{In the first version of \Refcite{Holguin:2023bjf}, we had employed \Herwig\texttt{7.2} compared to version \texttt{7.3} in this manuscript. We found a markedly improved stability of \Herwigxx compared to the previous version. This is reflected in the value of the perturbative constant $C$ in \tab{Cvalues} (which for $R = 1.2$, our default choice, is in significantly better agreement with \Pythiaxx and \Vinciaxx) as well as in the smallness of hadronization corrections displayed in \fig{had} below. For the sake of completeness, a more detailed comparison of versions \texttt{7.2} and \texttt{7.3} will be presented in the next revision of \Refcite{Holguin:2023bjf}.}, \Pythiaxx~\cite{Sjostrand:2014zea} and \Vinciaxx~\cite{vincia:2016} parton shower simulations, as well as the \Fastjet~\cite{Cacciari:2011ma} implementation of the anti-$k_T$ algorithm \cite{Cacciari:2008gp}. We focus on uncertainties in the calibration of the jet energy scale, which affects the reconstructed $p_{T,\mathrm{jet}}$, uncertainties in the constituent energy scale, which shift the momenta of the jet constituents, and different track reconstruction efficiency models that account for a $p_{T}$-dependent loss of particle tracks.
The effects of the various uncertainties are estimated by repeating the measurement of the ratio $\zeta_{t}/\zeta_W$ while changing the parameter characterizing an uncertainty source by one standard deviation.
The result is compared to the nominal analysis in each bin of $p_{T,\mathrm{jet}}$.
As a measure of the effect, we calculate the weighted average of the shifts in the extracted top quark mass for all $p_{T,\mathrm{jet}}$ bins.

To give realistic estimates of the uncertainties, we orient our models to the capabilities and existing calibrations of the CMS detector.
All simulations in this section have been performed using \Pythiaxx as the benchmark with the central \Pythia tune (CP5) as the default tune, similar to the simulated samples used in the CMS Collaboration~\cite{CMS:2019csb}.
The studies are obtained using particles at the hadron level without any detector simulation since the excellent angular resolution of the CMS detector in the order of mrad~\cite{CMS:2006myw} is not expected to have a considerable impact. However, precise uncertainties associated with unfolding, such as the impact of modelling in event generators, though expected to be small in track-based measurements, are considered beyond the scope of this work. This is because the unfolding uncertainties will strongly depend on the exact setup used in the measurement. However, the presented modelling variations can be interpreted as an upper limit of the expected impact in the unfolding.

Before continuing, a short word on the format and interpretation of the figures that present our results is needed.
In each figure, we show the extracted top mass parameter using \eq{C} as a function of $p_{T {\rm jet}}$. Red points show the output of \Pythiaxx using the CMS central \Pythia tune, CP5. This is used as a benchmark, while the other points show specific variations in the event generator modelling. When appropriate, a bottom panel is also attached, showing the deviation between the default setting and the associated variations. The uncertainties are nearly $100\%$ correlated between variations, so we do not show uncertainties on the bottom panel. Any shifts in the extracted top mass, due to variations in the event generator modelling, are expected to be nearly uniform over the moderate $p_{T {\rm jet}}$ range. Therefore, the shifts due to each variation averaged over $p_{T,\rm jet}$ bins are included in the legend with an uncertainty assigned from the standard deviation. These average shifts provide a good overall measure of the uncertainty due to a particular set of Monte Carlo variations.

\subsection{Jet Energy Scale}

\begin{figure}[t]
\centering
\includegraphics[height=0.46\textwidth]{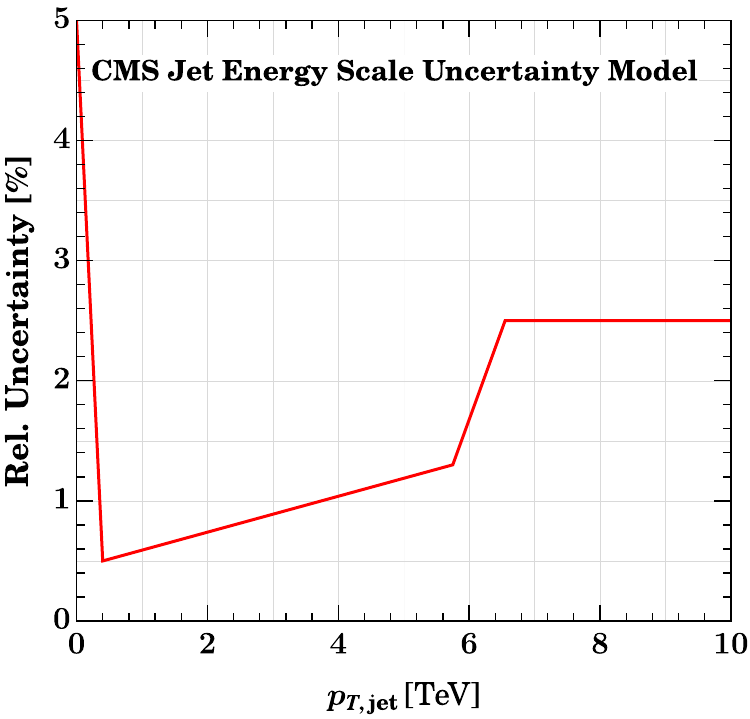}
\includegraphics[height=0.46\textwidth]{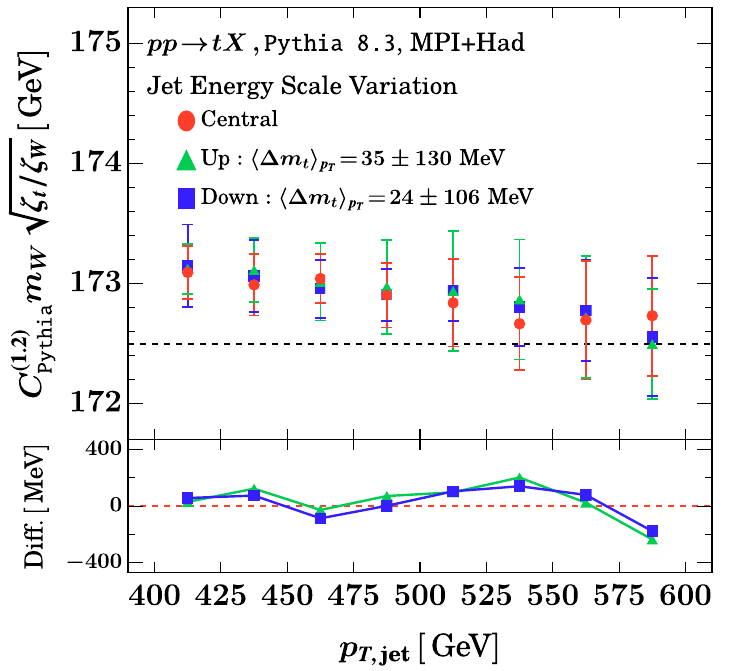}
\caption{Model of the relative CMS jet energy scale uncertainty as a function of the jet $p_T$~(left). Average shift in the top quark mass due to the jet energy scale uncertainty~(right). The shift is less than 100 MeV.}
\label{fig:jec}
\end{figure}

The finite knowledge of the jet energy scale in hadron collider experiments is crucial in most top quark mass measurements. It is the dominant uncertainty in direct extractions of $m_{t}$~\cite{CMS:2024irj,ATLAS:2018fwq,CMS:2024yqd,CMS:2023ebf} and also limits measurements of the jet mass in the boosted regime~\cite{CMS:2022kqg}, because it directly alters the observable used for the extraction of $m_{t}$.
Since the top quark $p_{T}$ changes the decay opening angle, both the peak location and the normalization of the correlator distribution, $\zeta_{t}$, are also affected by shifts of the jet energy scale.
An extraction of $m_{t}$ from the ratio of $\zeta_{t}$ and $\zeta_W$ however, should be robust against variations of $p_{T,\rm jet}$.

\begin{figure}[t]
\centering
\includegraphics[width=0.45\textwidth]{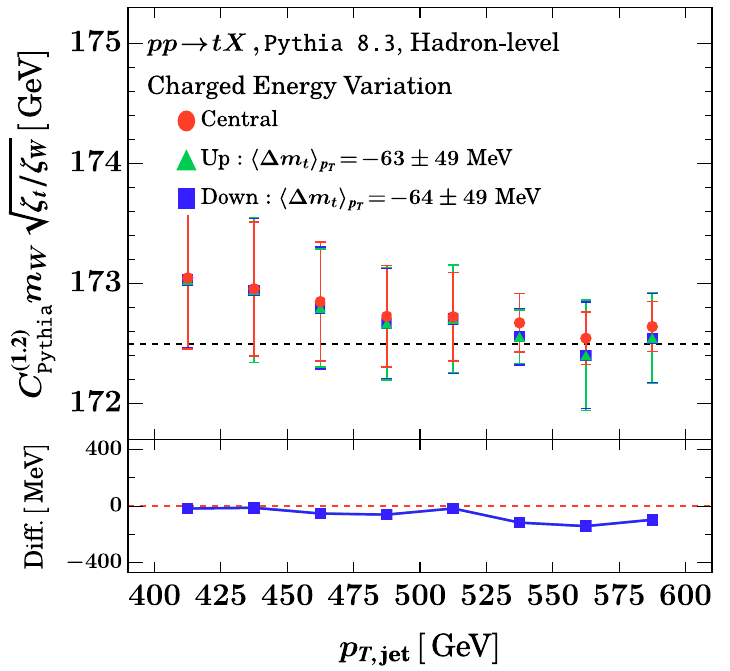}
\includegraphics[width=0.45\textwidth]{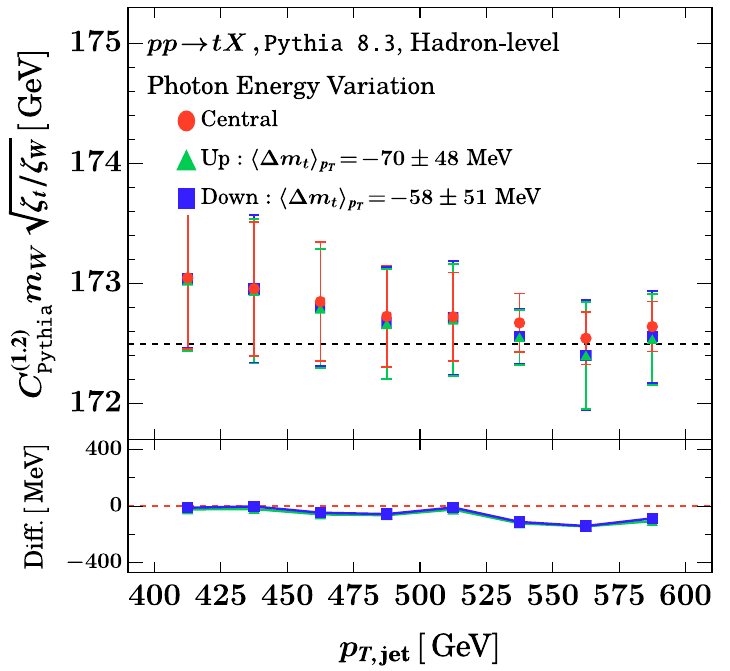}
\includegraphics[width=0.45\textwidth]{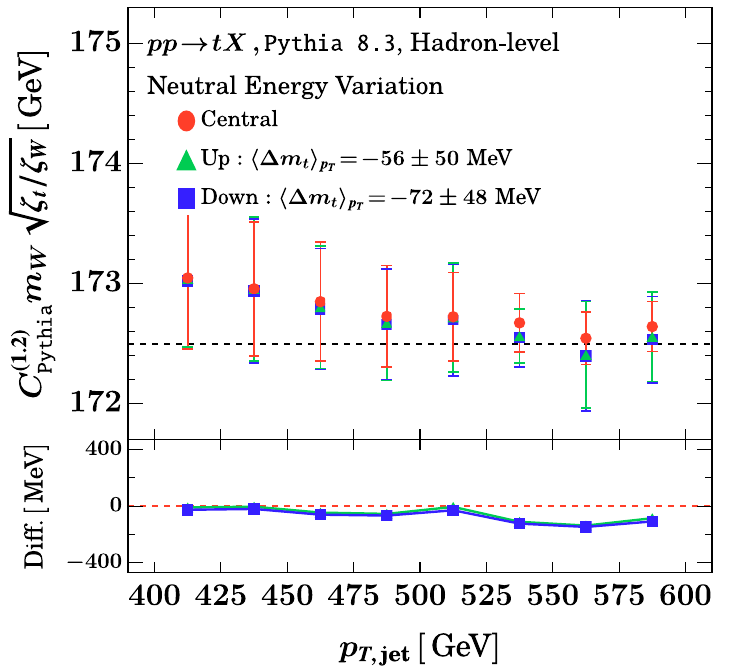}
\caption{The shift in the extracted top quark mass due to changes of the energy scale of charged jet constituents~(upper left), photons~(upper right), and other neutral constituents~(lower). The shift is found to be less than 100 MeV.}
\label{fig:cec}
\end{figure}

To test the presented measurement strategy against realistic variations, we model the CMS jet energy scale uncertainty~\cite{CMS:2016lmd,CMS-DP-2020-019} with the function displayed in \fig{jec}~(left).
Thus, the energy scale of most jets in this study is varied by 0.5--2.5\%.
We repeat the analysis while changing the jet momenta up and down according to the uncertainty defined above.
For this uncertainty, it was chosen to only change the jet momentum without altering its constituents, which are studied separately below.
The results are shown in \fig{jec}~(right).
Because of the ratio of the $\zeta_{t}$ and $\zeta_W$, the measurement is nearly unaffected by changes in the jet energy scale, which show a shift in the top quark mass below 100 MeV.

\subsection{Constituent Energy Scale}

In addition to uncertainties in the energy scale of the jet, we study the effect of varying the constituent momenta.
For this, we follow the strategy of an existing measurement using energy correlators inside jets by the CMS Collaboration~\cite{CMS:2024mlf} and separately vary the energy scale by 1\% for charged jet constituents, 3\% for photons, and 5\% for other neutral particles. The results are displayed in \fig{cec} and show negligible impact.

\begin{figure}
\centering
\includegraphics[width=0.47\textwidth]{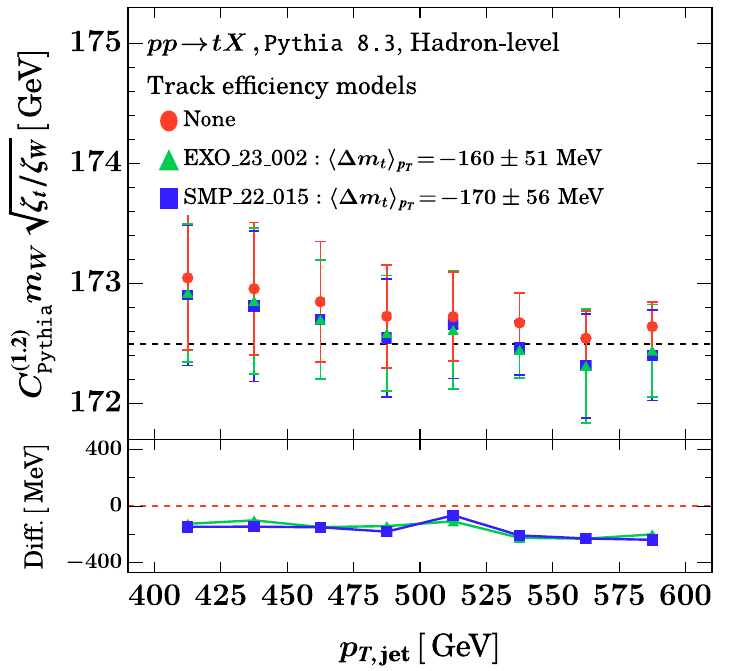}
\includegraphics[width=0.47\textwidth]{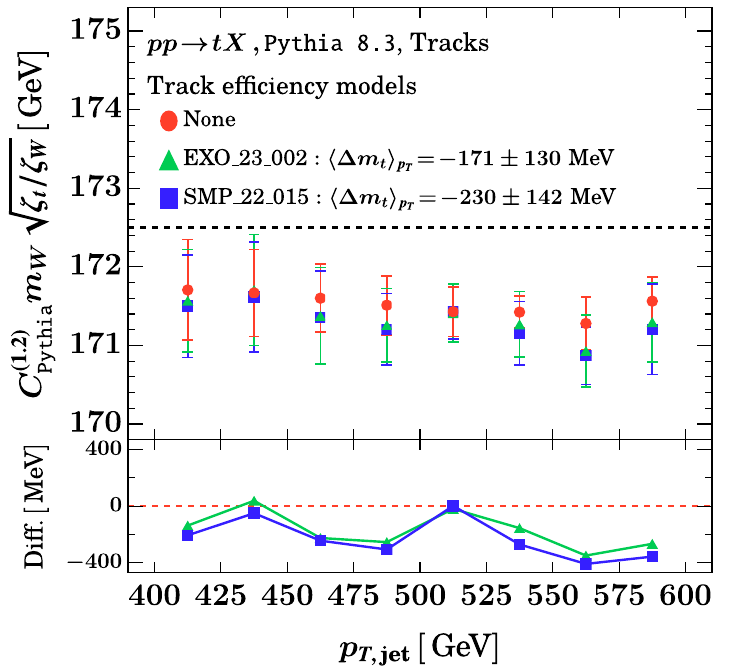}
\caption{The shift in the extracted top quark mass for two scenarios of track reconstruction inefficiencies for an analysis using all particles~(left) and only tracks~(right). In all cases, the observed shift is $\sim 200$ MeV. The horizontal dashed line shows the central value of the mass extracted at the parton level. }
\label{fig:tracks}
\end{figure}

\subsection{Tracking Efficiency}

We now focus on the feasibility of performing the proposed top mass measurement on charged particles (tracks) only. This is a unique advantage of the energy correlator measurement since other approaches to the top mass extraction from jet substructure suffer large corrections from the restriction to charged particle final states. In contrast, the restriction to tracks is a small effect to the energy correlator spectrum which can be systematically included in theoretical calculations \cite{Chang:2013rca,Chang:2013iba,Jaarsma:2023ell,Chen:2022pdu,Chen:2022muj,Jaarsma:2022kdd,Li:2021zcf}. This behavior is validated by \fig{tracks}, where the red circles in the right panel show the bin-by-bin top mass extraction performed on tracks. A remarkably small shift ($\lesssim 1~$GeV) relative to the central value of the all-hadrons result (the dashed line) is observed. This shift is computable to high precision from analytical theory, and it is exceedingly promising that only $10\%$ accuracy in the description of the nonperturbative track function input would be needed for this shift to be considered negligible.

\begin{figure}
\centering
\includegraphics[width=0.47\textwidth]{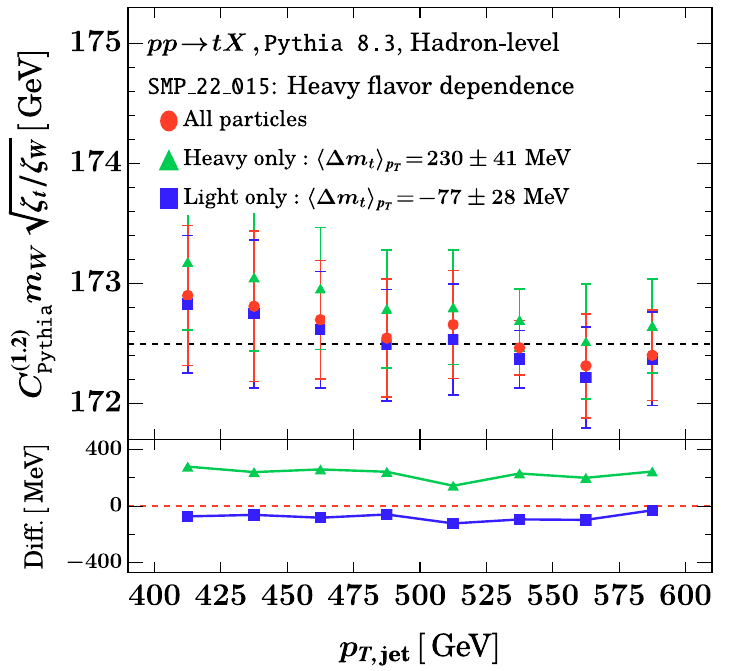}
\includegraphics[width=0.47\textwidth]{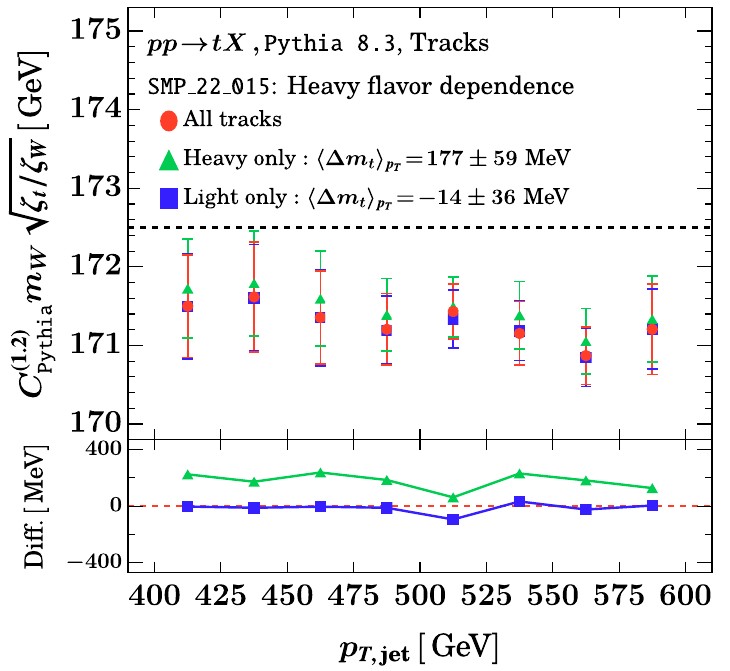}\\
\includegraphics[width=0.47\textwidth]{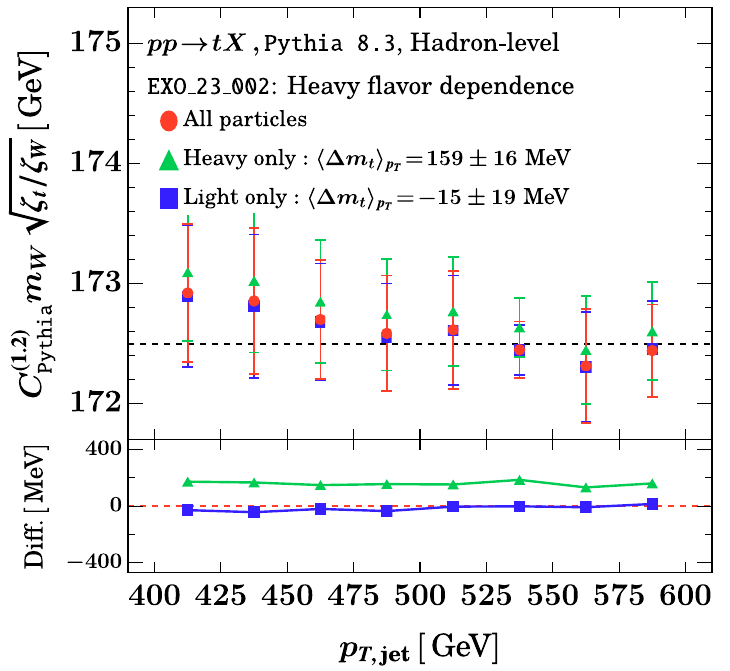}
\includegraphics[width=0.47\textwidth]{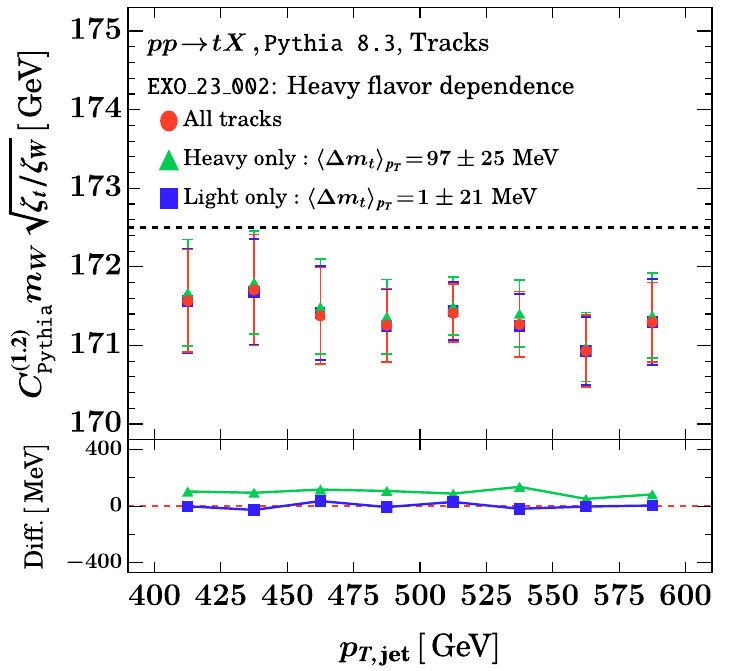}
\caption{The shift in the extracted top quark mass assuming two scenarios of track reconstruction applied only to particles from heavy quark decays or light quark decays. Results are compared for an analysis using all particles~(left column) and only tracks~(right column) as well as the models `\texttt{SMP-22-015}'~(upper row) and `\texttt{EXO-23-002}'~(lower row). The horizontal dashed line shows the central value of the mass extracted from an inclusive final state at the parton level. A remarkably small shift from the central value is observed when the measurement is restricted to charged particles. }
\label{fig:tracksHeavyLightSMP}
\end{figure}

While track-finding algorithms have a very high efficiency in current experiments, they can deteriorate in dense environments such as highly energetic jets.
In this context, we investigate two models, following the uncertainty estimations of two CMS analyses~\cite{CMS:2024mlf,CMS:2024nca}.
In model `\texttt{SMP-22-015}'~\cite{CMS:2024mlf} 3\% of all tracks within a jet are randomly excluded.
For model `\texttt{EXO-23-002}'~\cite{CMS:2024nca} a $p_T$ dependent approach is chosen such that 3\% of tracks with $p_\text{T} < 20\;\text{GeV}$ and 1\% of tracks with $p_\text{T} > 20\;\text{GeV}$ are randomly excluded.
The results are shown in \fig{tracks} for a version of the analysis with all particles~(left) and a version using only tracks~(right).
The observed shift due to different tracking efficiency models is small.

A known effect in detectors is the different jet response depending on the origin of a jet.
This difference in jet response between jets initiated by gluons, light quarks, and bottom quarks poses a serious challenge to many top quark mass measurements.
Because of this, calibrations of the jets used to reconstruct the top quark via the known $W$ boson mass peak suffer from large uncertainties.
Since in the measurement strategy discussed in this paper, the distribution associated with the $W$ is used comparably, we test the effect by applying the models `\texttt{SMP-22-015}' and `\texttt{EXO-23-002}' separately to particles that originate from a heavy flavor bottom quark or those that originate from a light flavor quark.
The results of the $m_{t}$ extraction are displayed in \fig{tracksHeavyLightSMP} for an analysis with all particles~(left) and a version using only tracks~(right). The effect remains small, particularly when the measurement is performed using only tracks.

\section{Sensitivity to subprocesses}
\label{sec:EventGen}

Having illustrated that our proposed measurement is minimally affected by experimental reconstruction uncertainties, it remains to be demonstrated that it is under theoretical control in the complicated hadron collider environment. The complete description of top quark production and decay at a hadron collider involves many subprocesses, as illustrated in \fig{fact}. From the theoretical perspective, a number of these subprocesses, such as color reconnection and the underlying event, are non-universal power corrections to factorization theorems. They are poorly understood and often require input from models. Therefore, to motivate the precision theoretical calculation of energy correlator observables on top quark decays in hadron colliders, it is desirable to show that these poorly understood nonperturbative contributions have a minimal effect on the energy correlator distribution. Furthermore, by studying the dependence of the energy correlator observable to these subprocesses, we can determine where theoretical effort must be concentrated to develop a precision first principles understanding of the proposed measurement.

In this section, we analyze the dependence of the energy correlator-based top quark measurement on physics subprocesses in the top quark production and decay. This is performed using parton shower generators, where the modelling of each subprocess can be modified. This discussion is split into \secn{NP} where we test the dependence on the modelling of event-sensitive nonperturbative effects. In \secn{sys_imp}, we then test the dependence on perturbatively improvable subprocesses.

\begin{figure}
\centering
\includegraphics[width=0.47\textwidth]{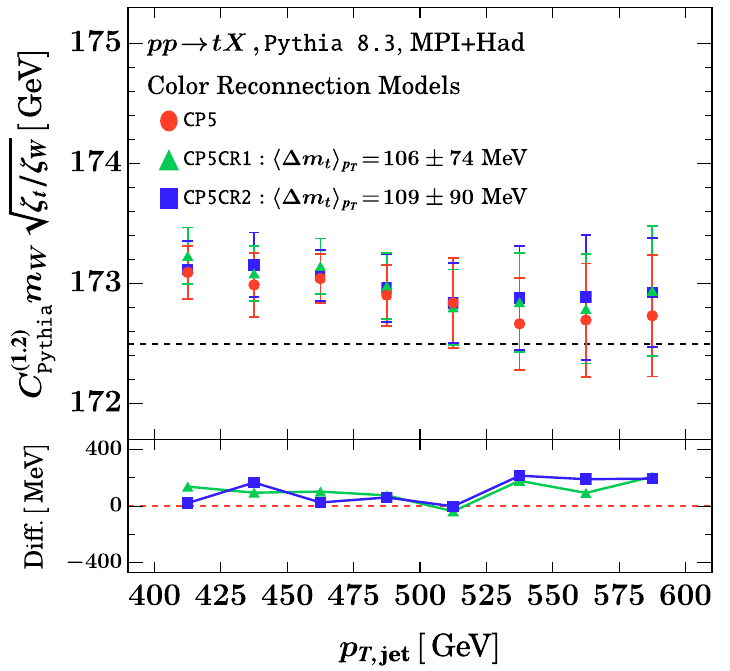}
\caption{The impact of the modelling of color reconnection on the extracted top mass is found to be less than 150 MeV.}
\label{fig:colorReconnection}
\end{figure}

In both subsections, we present variations of the event generator simulation of the top quark mass measurement and extract the associated shift in the top quark mass. Each variation is performed using \Pythiaxx as the benchmark with the central \Pythia tune (CP5) \cite{CMS:2019csb} as the default tune, consistently shown as red circles in the figures. Details of the tune can be found at \url{https://github.com/cms-sw/cmssw/blob/master/Configuration/Generator/python/MCTunes2017/PythiaCP5Settings_cfi.py}.

Throughout this section, we present figures in the style described in section~\ref{sec:exp}.

\subsection{Event Sensitive Nonperturbative Corrections}\label{sec:NP}

We begin by testing sensitivity to nonperturbative physics, where analytical control is very hard to achieve, usually requiring input from phenomenologically determined parameters and functions. This has been a limiting factor in previous approaches to top mass extraction from indirect measurements and in particular jet substructure \cite{Fleming:2007xt,Fleming:2007qr,Hoang:2017kmk,Hoang:2019ceu,Bachu:2020nqn}. Our energy correlator-based measurement has been designed to minimize sensitivity to these effects on theoretical grounds. Here, we will demonstrate that this is born out in simulation.

\subsubsection{Color Reconnection Modelling}

To evaluate the robustness against the modelling of color reconnection, we use two variations often considered in CMS studies of top quark mass measurements:
\begin{itemize}
\item QCD-inspired (CR1): \url{https://github.com/cms-sw/cmssw/blob/master/Configuration/Generator/python/MCTunes2017/PythiaCP5CR1TuneSettings_cfi.py}
\item Gluon move (CR2): \url{https://github.com/cms-sw/cmssw/blob/master/Configuration/Generator/python/MCTunes2017/PythiaCP5CR2TuneSettings_cfi.py}
\end{itemize}
Sensitivity to the color reconnection model, compared to the default, is shown in \fig{colorReconnection}. Variations of the order of $100$~MeV show that our observable is indeed robust to such effects.

\subsubsection{Underlying Event Tune}

\begin{figure}
\centering
\includegraphics[width=0.47\textwidth]{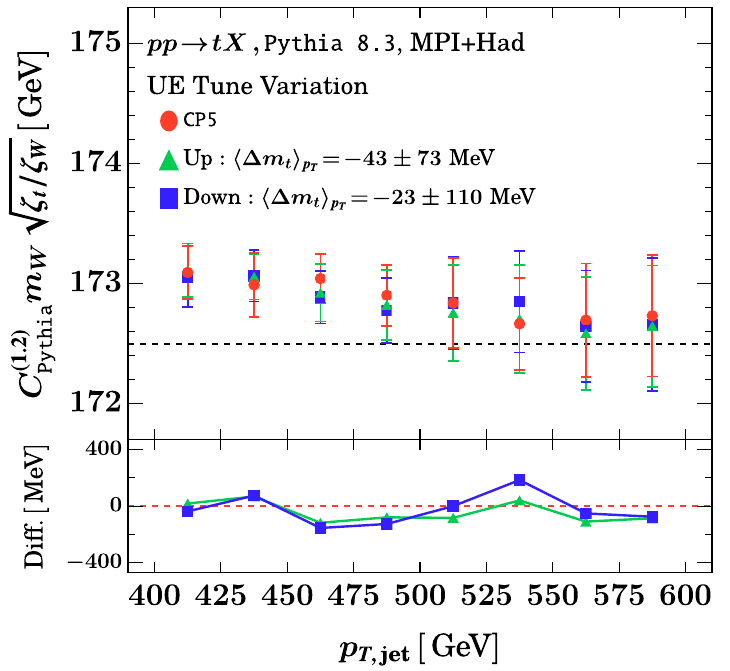}
\caption{The impacts of variations in the underlying event tune on the extracted top mass are found to be less than 100 MeV.}
\label{fig:ue}
\end{figure}

We also study the robustness of the modelling of the underlying event tune. We consider two variations, whose description can be found at:
\begin{itemize}
\item Up: \url{https://github.com/cms-sw/cmssw/blob/master/Configuration/Generator/python/MCTunes2017/PythiaCP5TuneUpSettings_cfi.py}
\item Down: \url{https://github.com/cms-sw/cmssw/blob/master/Configuration/Generator/python/MCTunes2017/PythiaCP5TuneDownSettings_cfi.py}
\end{itemize}
The results of the variation of the underlying event tune are shown in \fig{ue} and are found to impact the top quark mass extraction from energy correlators by less than 200 MeV.

\subsection{Sensitivity to Systematically Improvable Subprocesses}\label{sec:sys_imp}

Here, we focus on the sensitivity to the simulation of physical subprocesses relevant to the proposed measurement, wherein analytical computations with high accuracy are possible. We will show that under variation of the modelling of these processes, the top mass extraction changes only within simple trends, which can be accounted for by perturbation theory. These results strongly suggest that analytical computations from factorization theorems are possible with high precision.

\begin{figure}
\centering
\includegraphics[width=0.47\textwidth]{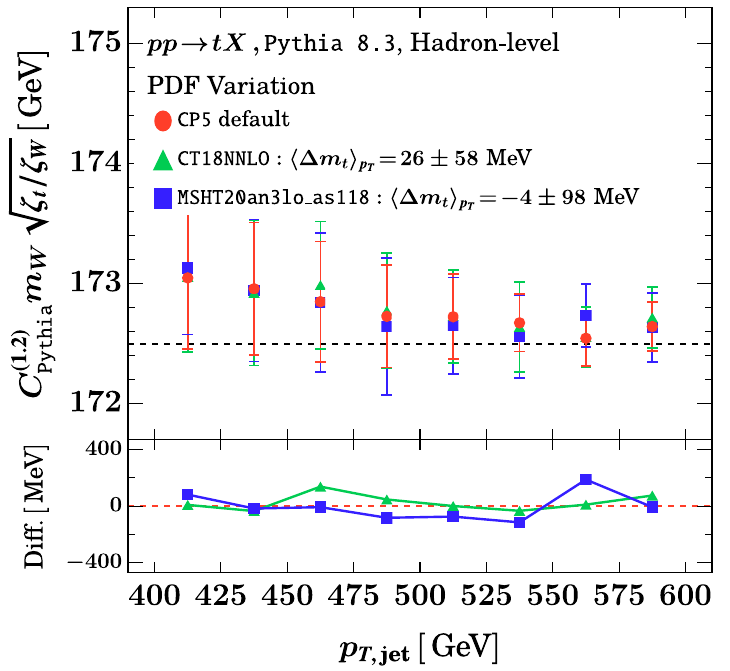}
\caption{Impact of variations of the PDFs are found to be minimal, showing that the observable isolates properties of the top quark decay.}
\label{fig:pdf}
\end{figure}

\begin{figure}
\centering
\includegraphics[width=0.48\textwidth]{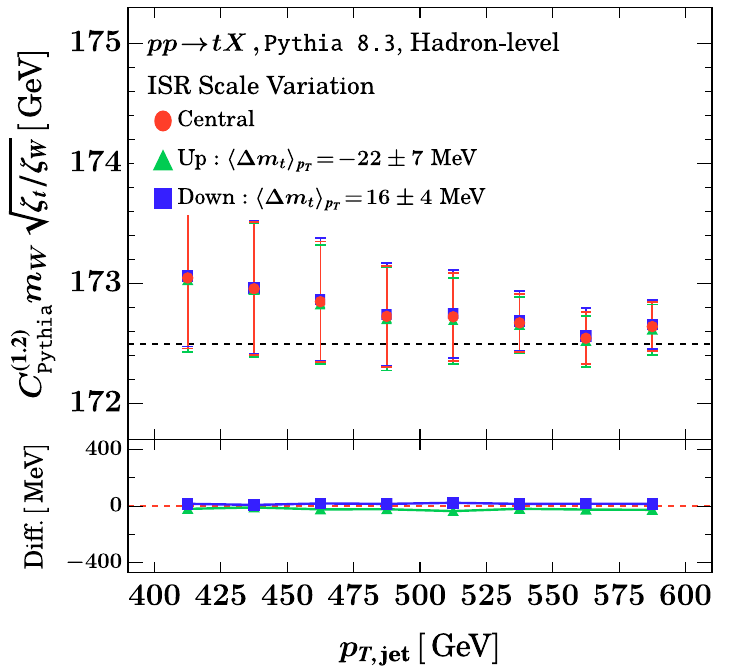}
\includegraphics[width=0.48\textwidth]{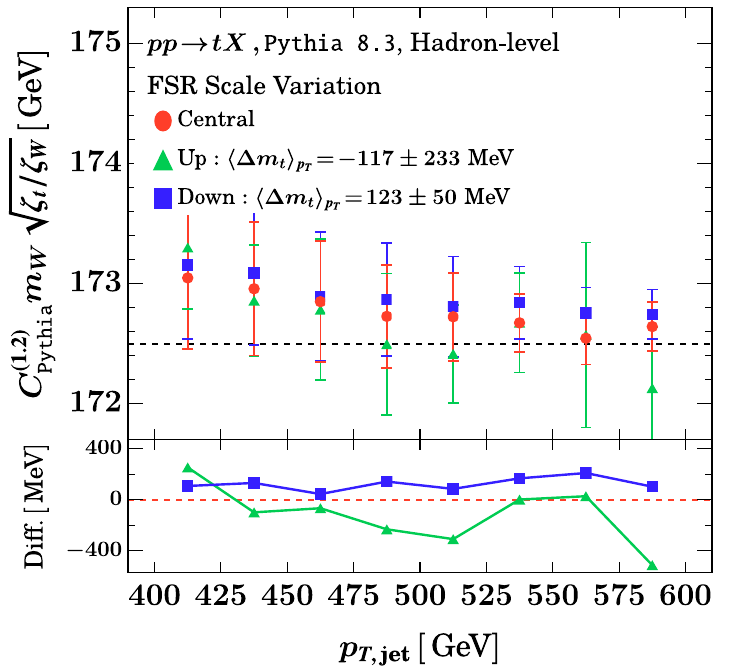}
\caption{Impact of variations of the ISR (left) and FSR (right) scales.}
\label{fig:fsr}
\end{figure}

We study the dependence on subprocesses moving systematically from the parton distribution functions and the description of the initial state to hadronization effects in the top quark decay. The goal is to show that the energy correlator-based observable is most sensitive to the perturbative description of the top quark decay, focusing theoretical activity on improving this aspect of the process and showing that it is robust to modelling uncertainties often studied in top quark mass extraction.

\subsubsection{Parton Distribution Functions}

We begin our discussion with the parton distribution functions (PDFs). While PDFs are under good theoretical control, separating the top quark mass extraction from PDF fitting is desirable. As highlighted in \Refcite{Holguin:2023bjf}, due to the use of the $W$ boson to calibrate the measurement, as well as the use of ratio observables, we expect that our energy correlator-based observable should be essentially insensitive to PDF variations.

In \fig{pdf}, we show the results of the PDF variations, which were found to have no significant impact.

\subsubsection{Initial and Final State Radiation}

We also study variations in the factorization scale for initial and final state radiation to further probe the dependence on different physical subprocesses. Again, while these are subprocesses under theoretical control, we wish to show that we are largely insensitive to them and that our observable isolates aspects of the top quark decay.

In CMS studies of top quark mass extractions, sensitivity to ISR/FSR is typically performed by varying the \Pythia parameters ``\texttt{isr:muRfac}'' and ``\texttt{fsr:muRfac}'' between 0.5 and 2, leading to a total of four variations.

These variations are shown in \fig{fsr}. Both variations were found to be not significant, however, notably, the effects were more prominent for FSR ($\approx 100$~MeV). This is expected since varying FSR modified the final state perturbative description of the top decay, to which our observable is sensitive by design.

\subsubsection{Matching Corrections to Stable Top Production}

\begin{figure}
\centering
\includegraphics[width=0.47\textwidth]{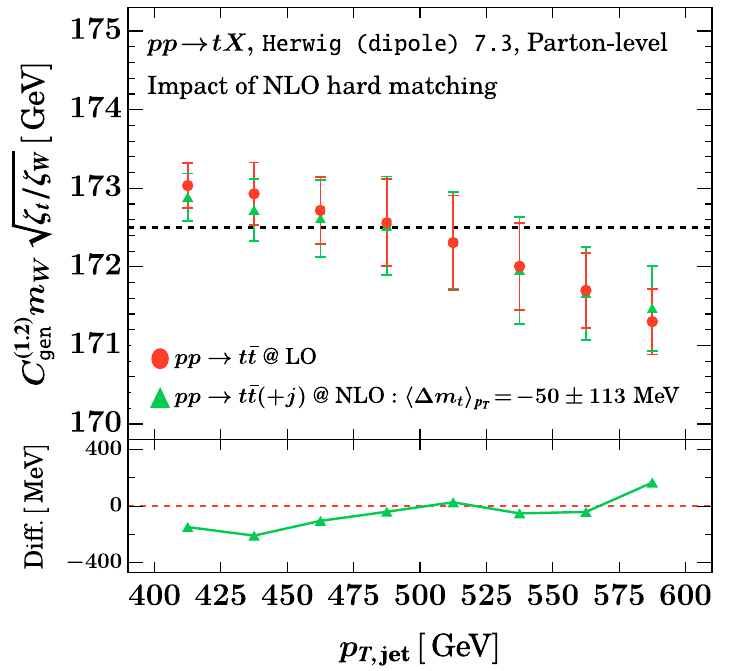}
\caption{Impact of NLO corrections to the stable top quark production process.}
\label{fig:nlo}
\end{figure}

We now focus on the impact of next-to-leading order (NLO) matching to the hard-process production of stable top quarks. Physically, this matching procedure modifies only the amount of gluonic radiation from the hard process before the top is produced and, therefore, the overall boost applied to the produced top quark. However, the boost of the top quark and the boost of the $W$ boson are strongly correlated, which is why the jet $p_T$ dependence cancels in the ratio of their peaks. Therefore, the extracted top mass is expected to be very stable under variation of the NLO matching. This is confirmed in our analyses.
Results are shown in \fig{nlo}. Including NLO corrections to the hard scattering has no significant impact. Again, this focuses the theoretical attention on improving the description of the top quark decay.

\subsubsection{Jet Radius}

In this subsection, we look at the response of the top mass extraction to different choices of the jet radius. Varying the jet radius will change the amount of the internal perturbative structure in the identified jet (i.e., whether or not perturbative radiation is lost). We therefore expect a simple perturbative dependence on $R$, which can be absorbed into $C^{R}_{\rm gen.}$. However, varying the jet radius also changes the nonperturbative properties of the jet, both in terms of the amount of underlying event contamination and the loss of particles due to fragmentation (each following competing geometric scaling laws \cite{Dasgupta:2007wa}). Our results show that nonperturbative effects from varying $R$ do not significantly modify the extracted top mass and that perturbative effects can be simply absorbed into the calculable factor $C^{R}_{\rm gen.}$.

\begin{figure}
\centering
\includegraphics[width=0.45\textwidth]{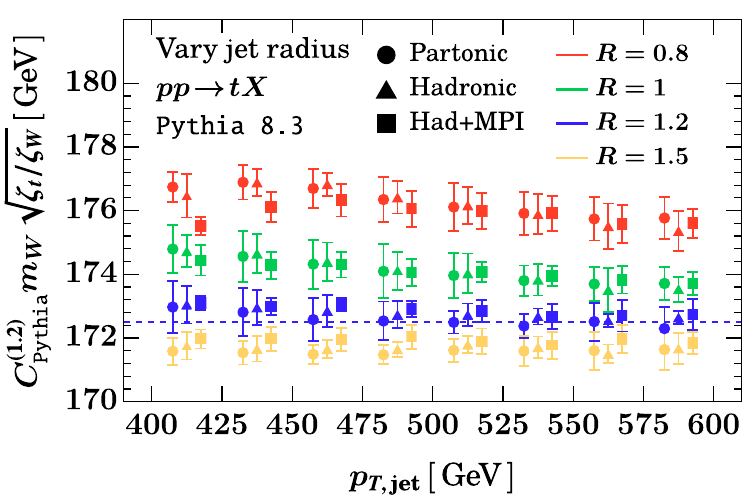}
\includegraphics[width=0.482\textwidth]{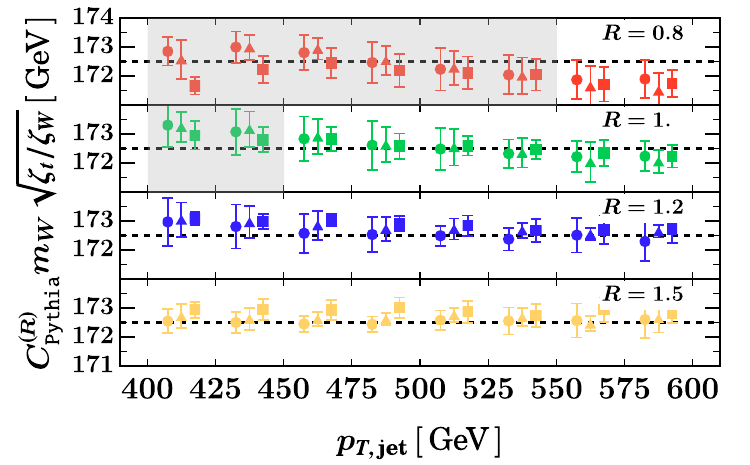}
\caption{Effects of the variation of the top jet radius. The grey hashed region illustrates the region, $R \lesssim \sqrt{2 \pi} \, m_{t}/p_{T,\mathrm{jet}}$, where jet radius effects are non-negligible.}
\label{fig:jetR}
\end{figure}

In \fig{jetR}, we show variations for different values of the top jet radius. For very small jet radii such as $R = 0.8$, the cone size is insufficient for lower $p_{T,\mathrm{jet}}$ bins. This can be seen as a significant sensitivity of the ratio to the underlying event contamination and a mild $p_{T,\mathrm{jet}}$ dependence. However, for larger jet radii $R > 1$, we find excellent robustness against hadronization and MPI effects, as well as a flat dependence on $p_{T,\mathrm{jet}}$.
For $R = 1.2$ and $R = 1.5$, we find a shift from hadronization and MPI, averaged over all $p_{T,\mathrm{jet}}$ values shown, to be
\begin{align}
&\text{$R$ = 1.2:}&
&\big(\Delta m_t\big)_{\rm had} = 201 \pm 109 \, \text{MeV} \, ,&
&\big(\Delta m_t\big)_{\rm MPI} = 99 \pm 72 \, \text{MeV} \, ,&\\
&\text{$R$ = 1.5:}&
&\big(\Delta m_t\big)_{\rm had} = 108 \pm 97 \, \text{MeV} \, ,&
&\big(\Delta m_t\big)_{\rm MPI} = 254 \pm 143 \, \text{MeV} \, ,&
\end{align}
where the uncertainty is statistical.
The common trend displayed in \fig{jetR} as well as the compatibility of the shifts for the different values of $R$ together with the agreement of the results for different values of the perturbative coefficient $C$ demonstrates that the dependence of the ratio observable on $R$ can be captured entirely in perturbation theory.

\subsubsection{Perturbative Top Decay and Showering}
\label{sec:varyingpertmodels}

The theoretical properties of the energy correlators motivate the expectation that our measurement of the top mass will be crucially sensitive to the perturbative description of the top decay. In this subsection, we will vary the event generator modelling of processes that approximate this perturbative input: in particular, the parton shower models, which at lowest order approximate the NLO top decay phase-space and matrix elements, and the top jet recoil schemes, which at the lowest order model the NLO effects from momentum conservation. Due to the complex and interwoven nature of the physics in event generator simulations, varying these perturbative inputs has knock-on effects throughout the rest of the perturbative and nonperturbative modelling. Nevertheless, our results demonstrate that only the perturbative components to these variations, produced in parton-level simulations, significantly affect the top mass extraction.

\begin{figure}
\centering
\includegraphics[width=0.48\textwidth]{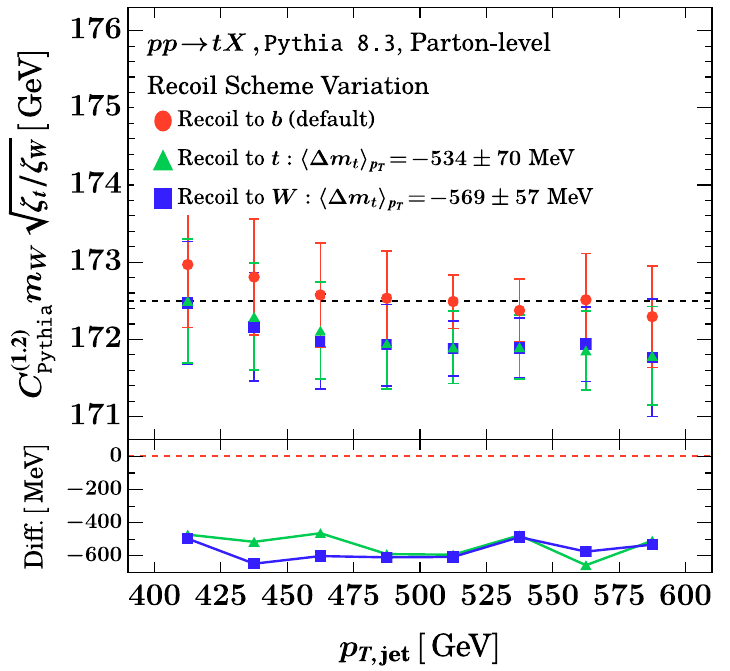}
\includegraphics[width=0.48\textwidth]{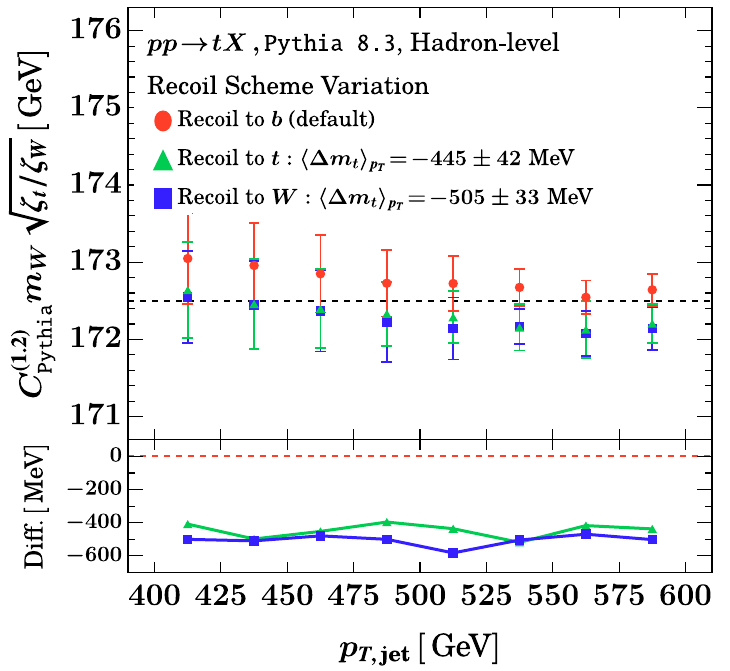}
\caption{Impact of the choice of recoil scheme in the top quark decay.}
\label{fig:recoil}
\end{figure}

The sensitivities to the extracted top mass are shown in \fig{recoil} for differing recoil schemes. A $500~$MeV shift is observed, which, as explained above, is of purely perturbative origin. \Tab{Cvalues} shows the change in the parameter $C$ with different perturbative parton shower models. Notably, the three dipole showers, which each handle the NLO top decay phase-space in an equivalent fashion, are consistent within $\pm 0.5\%$. In contrast, the Herwig angular ordered parton shower does not fill the complete NLO phase space and computes a value of $C$ larger by about $2\%$. This is consistent with the expectation from factorization, which is that underestimating the NLO contribution will lead to an overall smaller characteristic decay angle for the top quark.

\begin{figure}
\centering
\includegraphics[width=0.47\textwidth]{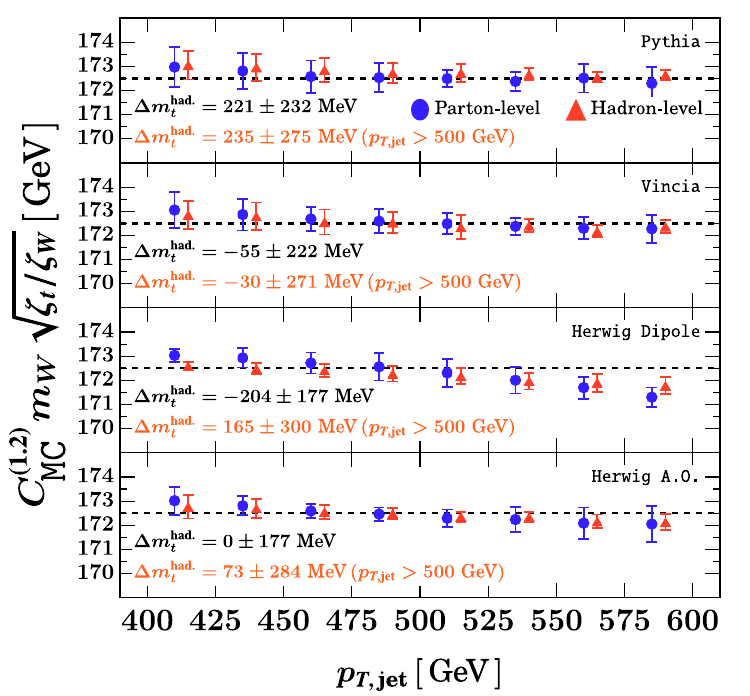}
\caption{Impact of hadronization corrections in different parton shower generators.}
\label{fig:had}
\end{figure}

\subsubsection{Hadronization}

Hadronization corrections have both universal and non-universal components. We have chosen to discuss the corrections from hadronization in this section since they are improvable, and indeed, there has been significant recent progress in understanding nonperturbative corrections to the energy correlators. However, much less is understood for the more complicated case of the three-point correlator appearing in our observable. It is, therefore, desirable if the nonperturbative corrections from hadronization are small.

In \fig{had}, we show the impact of hadronization corrections for different parton shower generators and for various values of $p_{T,\text{jet}}$. The observable has a minimal sensitivity of order $100~$MeV to hadronization corrections. This suggests that nonperturbative power corrections are under good control and confirms the results presented in \Refcite{Holguin:2023bjf}.

\subsubsection{$b$-fragmentation}

\begin{figure}
\centering
\includegraphics[width=0.47\textwidth]{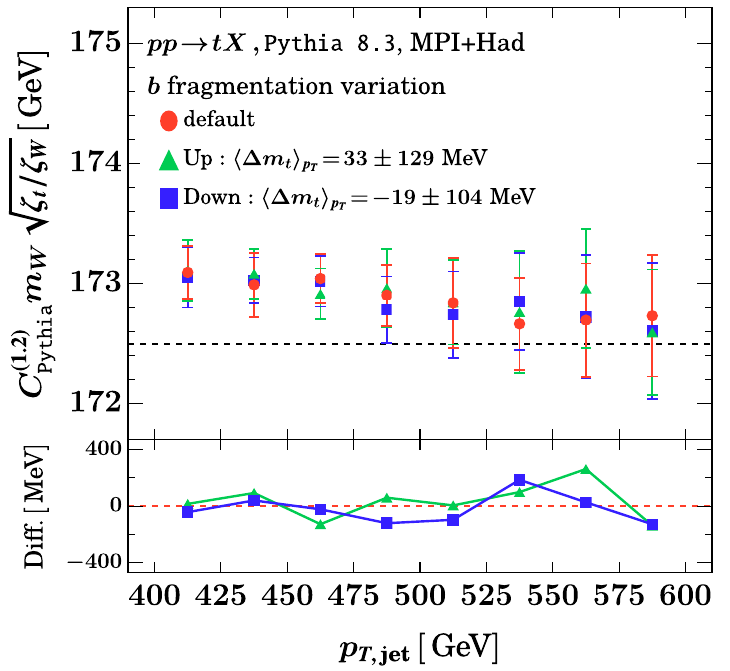}
\caption{Impact of variations in the parameterization of the $b$-fragmentation function.}
\label{fig:bfrag}
\end{figure}

Finally, we test the dependence on the modelling of the b-quark fragmentation function. Since the energy correlator is infrared and collinear safe, the $b$-quark fragmentation will enter as a universal power correction in the description of our observable. However, due to the specific design of our observable, we expect this power correction to be minimal.

Sensitivity to the b-quark fragmentation in top quark mass extraction studies is often obtained by varying the
\Pythia parameter \texttt{StringZ:rFactB}. From measurements at LEP, this parameter is tuned to a best-fit value of \texttt{StringZ:rFactB} = 1.056. In studies at CMS, it is varied between \texttt{StringZ:rFactB} = 0.855 and \texttt{StringZ:rFactB} = 1.252 to test the impact. The results of this variation are shown in \fig{bfrag}, where negligible impact is found. This illustrates insensitivity to the modelling of the b-quark fragmentation process.

\section{Conclusions}
\label{sec:conc}

In this paper, we have performed a systematic study of the recently proposed energy correlator-based top quark mass extraction~\cite{Holguin:2023bjf}, with a focus on characterizing experimental reconstruction uncertainties, and the dependence on different physics subprocesses in the top quark production and decay. Our analysis was performed using event generator simulations. However, we emphasize that event generators only play a role in our analysis as a proxy for future calculations based on first principles factorization theorems. Our analysis is designed to motivate theoretical investigation, experimental feasibility studies, and the eventual measurement of this observable.

On the experimental side, we have found minimal sensitivity to the jet energy scale resolution, constituent energy scale, and tracking efficiency, which are often the dominant uncertainties for jet substructure observables at hadron colliders. This motivates a more detailed exploration of the proposed observable at the detector level and the development of the necessary techniques for unfolding the three-point energy correlator on top quark decays.

On the theoretical side, we have shown minimal sensitivity to poorly understood nonperturbative subprocesses in the top quark production and decay. Our studies find that the measurement is dominantly sensitive to the perturbative description of the top quark decay. This strongly motivates the development of factorization theorems describing energy correlator observables on top quark decays, building on \cite{Fleming:2007xt,Fleming:2007qr,Hoang:2017kmk,Hoang:2019ceu,Bachu:2020nqn}, as well as a focus on improving the description of top quark decays. There has been significant recent progress in the perturbative description of higher point energy correlators \cite{Chen:2019bpb,Yan:2022cye,Yang:2022tgm,Yang:2024gcn,Chicherin:2024ifn}, and in higher order perturbative resummation in the back-to-back limit relevant for the calculation of energy correlators on top quark decays \cite{Moult:2018jzp,Gao:2023ivm,Gao:2019ojf,Ebert:2020sfi,Li:2016ctv,Duhr:2022yyp,Moult:2022xzt}. The inclusive top quark decay has been computed at NNLO \cite{Gao:2012ja}, and it will be essential to extend this to differential calculations of the energy correlator. It will also be important to improve the description of the top quark decay in publicly available parton shower generators, which is currently only described at leading order.

Our analysis emphasizes the energy correlator's unique robustness against event generator modelling and the modelling of detector acceptances. It highlights the potential for a successful measurement performed on charged particles (tracks) with improved angular resolution. The results of this paper pave the way for a precision determination of the top quark mass using energy correlators in a well-defined short-distance scheme with robust uncertainty estimates.

\acknowledgments
We thank Matt Leblanc, Jennifer Roloff, Meng Xiao, and HuaXing Zhu for useful discussions. We thank Simon Pl\"atzer for help with {\tt Herwig} simulations. J.H. is supported by the Leverhulme Trust as an Early Career Fellow. I.M. is supported by the DOE grant HEP GR120647. D.S. is supported by the Austrian Science Fund~(FWF, grant P33771). The computational results were obtained using the Vienna Bio Center and the Austrian Academy of Sciences CLIP cluster at~\url{https://www.clip.science/}. The authors would like to express special thanks to the Mainz Institute for Theoretical Physics (MITP) of the Cluster of Excellence PRISMA$^+$ (Project ID 390831469) for its hospitality and support.

\appendix

\bibliography{spinning_gluon}

\end{document}